\documentclass[aps, rmp, twocolumn, amssymb, groupedaddress, floatfix,nofootinbib]{revtex4}

\pdfoutput=1
\usepackage{amsmath}
\usepackage{amsfonts}
\usepackage{graphicx}

\usepackage{multirow}
\usepackage{textcomp}
\usepackage{array}

\newcommand{\mon}{\begin{displaymath}}
\newcommand{\moff}{\end{displaymath}}

\newcommand{\eon}{\begin{equation}}
\newcommand{\eoff}{\end{equation}}
\newcommand{\eaon}{\begin{eqnarray}}
\newcommand{\eaoff}{\end{eqnarray}}

\begin{document}

\title{Pleiotropic consequences of adaptation across gradations of environmental stress in budding yeast}

\author{Elizabeth R. Jerison$^{1,2}$}
\author{Sergey Kryazhimskiy$^{2,3}$}
\author{Michael M. Desai$^{1,2,3}$}
\affiliation{\mbox{${}^1$Department of Physics}}
\affiliation{\mbox{${}^2$FAS Center for Systems Biology}}
\affiliation{\mbox{${}^3$Department of Organsimic and Evolutionary Biology, Harvard University, Cambridge MA 02138}}

\begin{abstract}

Adaptation to one environment often results in fitness gains and losses in other conditions. To characterize how these consequences of adaptation depend on the physical similarity between environments, we evolved 180 populations of {\em Saccharomyces cerevisiae} at different degrees of stress induced by either salt, temperature, pH, or glucose depletion. We measure how the fitness of clones adapted to each environment depends on the intensity of the corresponding type of stress. We find that clones evolved in a given type and intensity of stress tend to gain fitness in other similar intensities of that stress, and lose progressively more fitness in more physically dissimilar environments. These fitness trade-offs are asymmetric: adaptation to permissive conditions incurred a smaller trade-off in stressful conditions than vice versa. We also find that fitnesses of clones are highly correlated across similar intensities of stress, but these correlations decay towards zero in more dissimilar environments. To interpret these results, we introduce the concept of a joint distribution of fitness effects of new mutations in multiple environments (the JDFE), which describes the probability that a given mutation has particular fitness effects in some set of conditions. We find that our observations are consistent with JDFEs that are highly correlated between physically similar environments, and that become less correlated and asymmetric as the environments become more dissimilar. The JDFE provides a framework for quantifying evolutionary similarity between conditions, and forms a useful basis for theoretical work aimed at predicting the outcomes of evolution in fluctuating environments.

\end{abstract}

\date{\today}
\maketitle

\section*{Introduction}

Adaptation to one environment often leads to costs in other conditions. These costs, which can arise from the degradation of unused functions or from direct trade-offs between traits, play a prominent role in evolutionary theories of adaptive diversification \citep{orr2000adaptation,RaineyTravisano1998} and speciation \citep{coyne2004speciation}, and are a prerequisite for the evolution of complex generalist strategies such as regulation \citep{dekel2005optimality}. This has motivated much previous effort in laboratory evolution experiments to find and characterize fitness tradeoffs between distinct environments \citep{bennett2007experimental,turner2000cost,MacLeanBellRainey2004,DuffyTurnerBurch2006,Bell2010,hietpas2013shifting,JasminZeyl2013,leiby2014metabolic}. Numerous cases have been analyzed --- for example, experiments in phage have shown that specialization to a novel host can lead to tradeoffs on the ancestral host \citep{turner2000cost,DuffyTurnerBurch2006}, and experimental adaptation of \emph{E. coli }to low temperature can lead to tradeoffs at high temperature \citep{bennett2007experimental}.

While tradeoffs have been observed between certain conditions, adaptation to a one environment can also sometimes confer an advantage in others \citep{BennettLenski1993, leiby2014metabolic, wenger2011hunger}. For example, recent work shows that long-term laboratory adaptation of \emph{E. coli }to low-glucose media confers advantages on other carbon sources \citep{leiby2014metabolic}. Qualitatively similar results have been observed in budding yeast \citep{wenger2011hunger}. In general, we expect that adaptation to one condition will yield fitness benefits in other conditions that are ``similar'' in some evolutionary sense, and will incur fitness costs in more dissimilar conditions. However, the transition between correlated adaptation to similar environments and tradeoffs across sufficiently different conditions remains poorly understood.

The most direct approach to characterize these correlations and tradeoffs is to study the raw material on which evolution acts: individual mutations. Along these lines, several earlier studies have screened large collections of mutants in a range of laboratory environments \citep{qian2012genomic, bank2014bayesian, hietpas2013shifting, gerstein2012parallel, ehrenreich2010dissection, Jasnos2008, wang2014sensitivity, smith2008gene}. These studies find numerous individual mutations that exhibit fitness tradeoffs, and others that confer advantages or disadvantages across multiple conditions. However, this approach inevitably involves a biased subset of mutations, which makes it difficult to use these results to predict evolutionary outcomes across environments.

Rather than studying individual mutations, one can also directly measure patterns of fitness gain and loss after adaptation to different environments \citep{cooper2000population, MacLeanBellRainey2004, bataillon2011cost, leiby2014metabolic, rodriguez2014different}. Two measures of the evolutionary outcomes are of particular interest. First, the average change in fitness in one condition after evolution in another indicates whether a population as a whole would thrive or suffer if the environment were to switch. Second, the correlation between the fitnesses of evolved clones in one condition and their fitnesses in another condition informs us about which individuals within the population are more or less likely to survive after an environmental shift, and hence the extent to which fluctuations between environments will slow adaptation to either.

Here, we measure how these two quantities change as environmental conditions become increasingly dissimilar along a particular physical dimension. To do so, we evolved 180 replicate budding yeast populations across nine conditions that range from permissive to stressful, by tuning four physical variables (temperature, pH, osmotic stress, and glucose concentration). We analyze how adaptation to each condition leads to correlated adaptation or tradeoffs across different degrees of the same type of stress. We address three main questions. First, how does the average fitness of a strain adapted to a particular stress change across different degrees of that stress? Second, how correlated are fitnesses in two environments as a function of their similarity? Is the correlation structure similar for different environmental variables, or qualitatively different? Finally, are the patterns of fitness gain or loss after adaptation to stressful and permissive conditions symmetric? If not, are there general patterns to the asymmetry?

Previous work analyzing tradeoffs between distinct environments has largely focused on understanding whether observed tradeoffs are driven by antagonistic pleiotropy, in which the same mutations that confer an advantage in one environment have a corresponding cost in another \citep{WagnerZhang2011}. Alternatively, these tradeoffs could be a side effect of mutation accumulation, in which neutral mutations accumulating in one environment have fitness costs in another. The distinction between antagonistic pleiotropy and mutation accumulation is important in understanding whether observed fitness declines are driven by selection due to inherent tradeoffs between traits, or are simply the consequence of the neutral degradation of unused functions \citep{rose1980test, MacLeanBellRainey2004, bennett2007experimental, leiby2014metabolic}.

To interpret our results, we generalize the ideas of antagonistic pleiotropy and mutation accumulation to the concept of a joint distribution of fitness effects of new mutations in multiple environments (the ``JDFE''). The JDFE can also be thought of as an extension of the idea of a distribution of fitness effects of new mutations: it describes the probability that a new mutation has specific fitness effects in each of several environments. Intuitively, the JDFE describes the underlying correlations between fitness effects of new mutations in multiple environments. Just as the DFE is a crucial parameter for predicting adaptation in a constant environment, the JDFE encapsulates the information about the available mutations that is important in predicting evolutionary outcomes in a fluctuating environment. In contrast to the discrete distinction between antagonistic pleiotropy and mutation accumulation, the JDFE provides the framework needed to interpret patterns of correlated adaptation to similar environments as well as tradeoffs across different conditions.

\section*{Methods}

\begin{table*}[!ht]
  \begin{tabular}{p{0.75cm}p{0.75cm}|>{\centering}p{1.25cm}>{\centering}p{1cm}>{\centering}p{1cm}>{\centering}p{1cm}|p{5cm}p{1cm}}
    \hline
    \multirow{2}{*}{Stress} &  \multirow{2}{*}{Level} & Glucose, & NaCl, & T, & \multirow{2}{*}{pH} & \centering Measurement  & \multirow{2}{*}{\centering Color}  \\
    & & \% & M & \textdegree C &  & \centering Environments & \\
    \hline
    None               &  Ctrl.$^*$& 2.00 & 0   & 30 & 5.0 & All           & \includegraphics{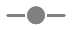} \\
    \hline
    \multirow{3}{*}{Salt}
                       & Low$^*$     & 2.00 & 0.2 & 30 & 5.0 & Ctrl, low, med, and high salt &  \includegraphics{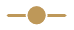} \\
                       & Med      & 2.00 & 0.4 & 30 & 5.0 & N/A           &  \\
                       & High$^*$    & 2.00 & 0.8 & 30 & 5.0 & Ctrl, low, med, and high salt &  \includegraphics{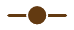}\\
    \hline
    \multirow{3}{*}{Temp}
                       & Low$^*$     & 2.00 & 0   & 21 & 5.0 & Ctrl, low, med, and high temp &  \includegraphics{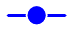} \\
                       & Med      & 2.00 & 0   & 34 & 5.0 & N/A           &  \\
                       & High$^*$    & 2.00 & 0   & 37 & 5.0 & Ctrl, low, med, and high temp &  \includegraphics{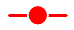} \\
    \hline
    \multirow{3}{*}{pH}
                       & Low$^*$     & 2.00 & 0   & 30 & 3.8 & Ctrl, low, med, and high pH &  \includegraphics{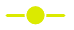} \\
                       & Med$^*$  & 2.00 & 0   & 30 & 6.0 & Ctrl, low, med, and high pH & \includegraphics{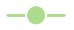} \\
                       & High$^*$    & 2.00 & 0   & 30 & 7.3 & Ctrl, low, med, and high pH & \includegraphics{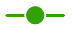} \\
    \hline
    \multirow{2}{*}{Glu}
                       & Low$^*$     & 0.07 & 0   & 30 & 5.0 & Ctrl, low, and med glu     &  \includegraphics{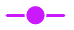} \\
                       & Med      & 0.40 & 0   & 30 & 5.0 & N/A     &  \\
    \hline
  \end{tabular}
\caption{{\bf Environments used in the experiment.} Environments in which we conducted evolution are marked with an asterisk; other environments were used only for fitness measurements. The fitness of each clone evolved in a given environment was assayed in a range of environments indicated in the column ``Measurement environments.''}
\label{tableone}
\end{table*}

\subsection*{Strains and Environments}

All strains used in this study were derived from yGIL104, a haploid yeast strain with a W303 background and the genotype \emph{URA3, leu2, trp1, CAN1, ade2, his3, bar1$\Delta$::ADE2, MATa} \citep{LangMurray2008}. yGIL104 was the ancestor of the evolution experiment. The fluorescently-labeled strain ySAK0449 used for competitive fitness assays was constructed from yGIL104 by integrating the {\em HIS3-ymCitrineM233I }cassette it into the {\em his3 }locus as described in Ref.~\citep{Kryazhimskiy2014}.

All media used in this study were based on Synthetic Complete (SC) Medium (2 g/L of SC, Sunrise Science catalog \#1300-030, 6.7 g/L of Yeast Nitrogen Base with Nitrogen, Sunrise Science catalog \#1501-100, and 20 g/L of glucose). To prevent bacterial contamination, all media also contained Ampicillin (100 $\mu$g/mL) and Tetracycline (25 $\mu$g/mL). Evolution in the `Control', `Low temperature', and `High temperature' environments was conducted in this SC medium. For `Low salt', `Medium salt', and `High salt'  environments, we supplemented SC with an additional 0.2 M, 0.4 M, or 0.8 M NaCl, respectively. For the pH-stress environments, we used SC media buffered with a citrate-phosphate system. For `Low pH' (pH 3.8), we supplemented SC with 35.4 mM disodium phosphate and 32.3 mM citric acid. For `Medium pH' (pH 6), we supplemented SC with 64.2 mM disodium phosphate and 17.9 mM citric acid. For `High pH' (pH 7.2), we supplemented SC with 80 mM disodium phosphate. For the `Low glucose' and `Medium glucose' environments, we reduced the amount of glucose in our SC media to 0.7 g/L and 4 g/L of glucose, respectively. For all but the temperature-stress environments, populations were incubated at 30\textdegree C. Populations evolved in the low, medium and high temperature environments were incubated at 21\textdegree C, 34\textdegree C, and 37\textdegree C, respectively. We summarize all of these environments in Table~\ref{tableone}.

\subsection*{Evolution}

We founded 20 replicate populations in each of nine environments: control, low and high salt, low and high temperature, low, medium, and high pH, and low glucose. Each population was founded by a single independent colony of yGIL104, picked from a YPD agar plate. All populations were grown in 96-well polystyrene plates (Corning, VWR catalog \#29445-154) without shaking. We maintained a set of blank wells in each plate to prevent plate misidentification and to control for cross-contamination. We observed only one contamination of a blank well during the course of the experiment.

We propagated all populations in batch culture. In each dilution cycle, cells were first resuspended using a Titramax 100 plate shaker, and the culture was diluted into fresh media using a Biomek FX pipetting robot. The dilutions and serial transfer schedules were chosen to keep the population bottleneck sizes constant across environments, to keep all populations on either a 24, 36 or 48 hour cycle, and so that populations would spend between 4 and 7 hours at saturation density. For the low glucose populations, this was not possible due to slow growth; these populations were therefore diluted to keep the population bottleneck similar to other environments, but did not reach saturation at the end of each dilution cycle. Populations from the control, low salt, high temperature, and low and medium pH environments were transferred on a 24 hour cycle. Populations from the high salt and low temperature environments were transferred on a 36 hour cycle. Finally, populations from high pH were transferred on a 48 hour cycle until cycle 26 (generation 234); thereafter, they were transferred on a 36 hour cycle. All populations were diluted $1:2^9$ each cycle, except for the populations in low glucose, which were diluted $1:2^6$ each cycle. The final population size at the end of each dilution cycle was $\sim 6\times10^6$ in the control, low temperature, low salt, low pH and medium pH environments. The final population size was $\sim 5\times10^6$ in high pH and high salt environments. The final population size at the end of the dilution cycle was $\sim 5\times10^5$ in low glucose. During the course of the experiment, three populations (one each from the low pH, medium pH, and low temperature environments) were lost due to pipetting error.

A single clone was picked from each evolved population at the following generations: generation 820 for control and low salt environments, generation 730 for the low glucose environment, generation 750 for low and medium pH environment, generation 500 for the high pH environment, generation 610 for low temperature and high salt environments. All fitness values plotted are scaled to be fitness gain/loss per 610 generations.

\begin{figure*}[!ht]
\begin{center}
\includegraphics{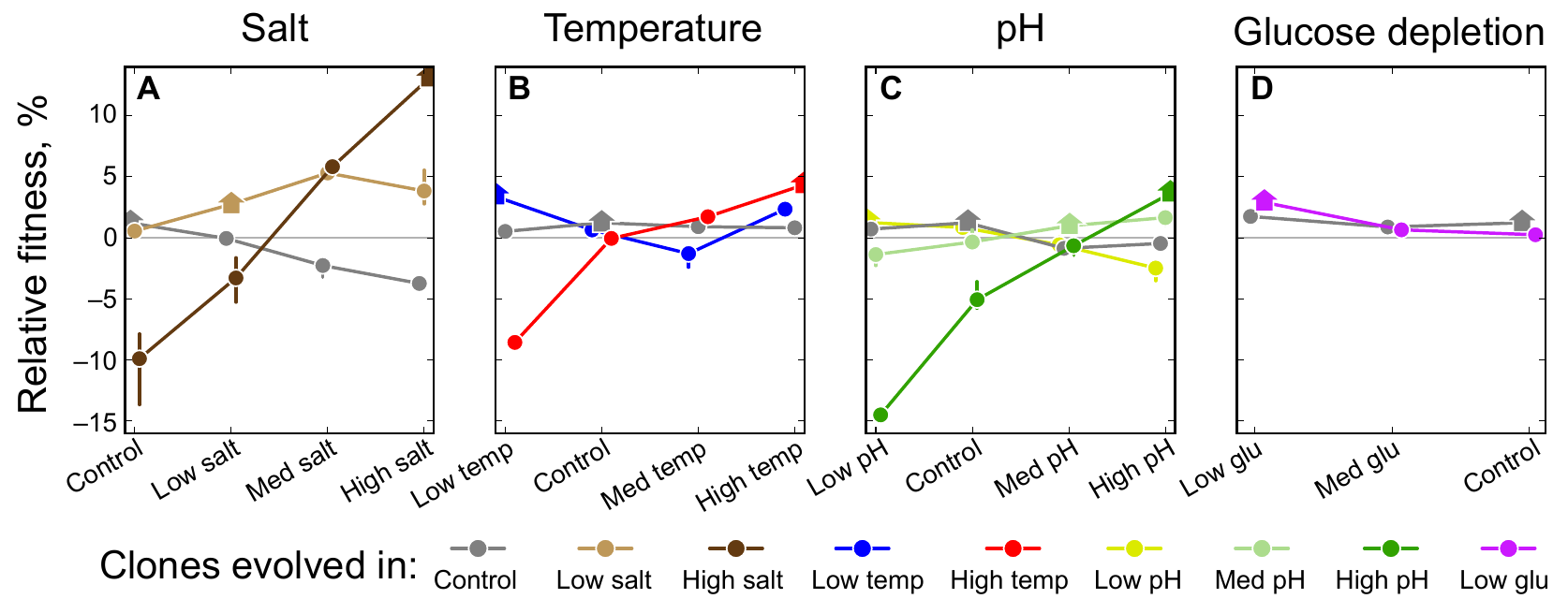}
\end{center}
\caption{{\bf Median fitness of a group of clones evolved in a particular condition, across a range of intensities of the same stress.} Colors correspond to each group's home environment (also indicated by home symbols) as indicated in the legend and in Table~\ref{tableone}. Error bars denote the interquartile range of the error distribution for the median as calculated by bootstrapped resampling (see Methods).}
\label{fig: median fits}
\end{figure*}

\subsection*{Fitness Assays}

Flow-cytometry-based competitive fitness assays were based on the protocol described in \citep{Kryazhimskiy2014}. For this study, this protocol was slightly modified as follows. All lines to be competed, and also the fluorescently-labeled reference strain, were diluted 1:32 into SC and grown for 24 hours at 30\textdegree C. All lines were then preconditioned in the assay environment: they were diluted with the appropriate dilution factor for that condition, and grown separately for one cycle, either 24 or 36 hours (see previous section). After this, the preconditioned reference and experimental lines were mixed in a 1:1 ratio. Frequencies of the fluorescent reference and evolved lines were measured using flow cytometry at the end of the first and third cycles after mixing (for low glucose, frequencies were measured at the first and fourth cycles after mixing).

Fitness was calculated as
\begin{equation*}
  s = \frac{1}{t_{\mathrm{assay}} } \ln \left(\frac{n_{\mathrm{final, evolved}} }{n_{\mathrm{final, ref}}} \frac{n_{\mathrm{init, ref}}}{ n_{\mathrm{init, evolved}}} \right),
\end{equation*}
where $n$ is the number of cells of a particular type, `ref' refers to the fluorescently-labeled ancestor (see above), `evolved' refers to clones chosen as described above, and $t_\mathrm{assay}$ is the time elapsed, in generations, between the initial and final FACS measurements. All fitness measurements were done in triplicate.

\subsection*{Analysis}

Fitness calculations and statistical analyses were done using custom python scripts, available upon request. The error bars on the median fitnesses across a group of clones in Figure~\ref{fig: median fits} and the correlations in Figure~\ref{fig: spearman all} show the interquartile range of the error distribution determined by bootstrapped resampling from the data. To resample the data, error models were built to describe the error between replicate fitness measurements of a single clone. Two error models were used: one for clones with fitnesses between $6\%$ and $-6\%$, and one for clones with fitnesses less than $-6\%$ or greater than $6\%$. The distinction was made because our assay is less precise for the latter class of clones, since one competitor tends to be at low frequency at the final timepoint of the assay. For each class of clones, we fit a Gaussian error model, using the average variance among replicate fitness measurements for all clones in that class. For the resampling, each clone was assigned its measured fitness, plus an error sampled from the appropriate error model. A set of twenty clones was then sampled with replacement, and the statistic of interest (either median fitness or Spearman correlation coefficient) was calculated on this resampled clone group.

\begin{figure*}[!ht]
\begin{center}
\includegraphics{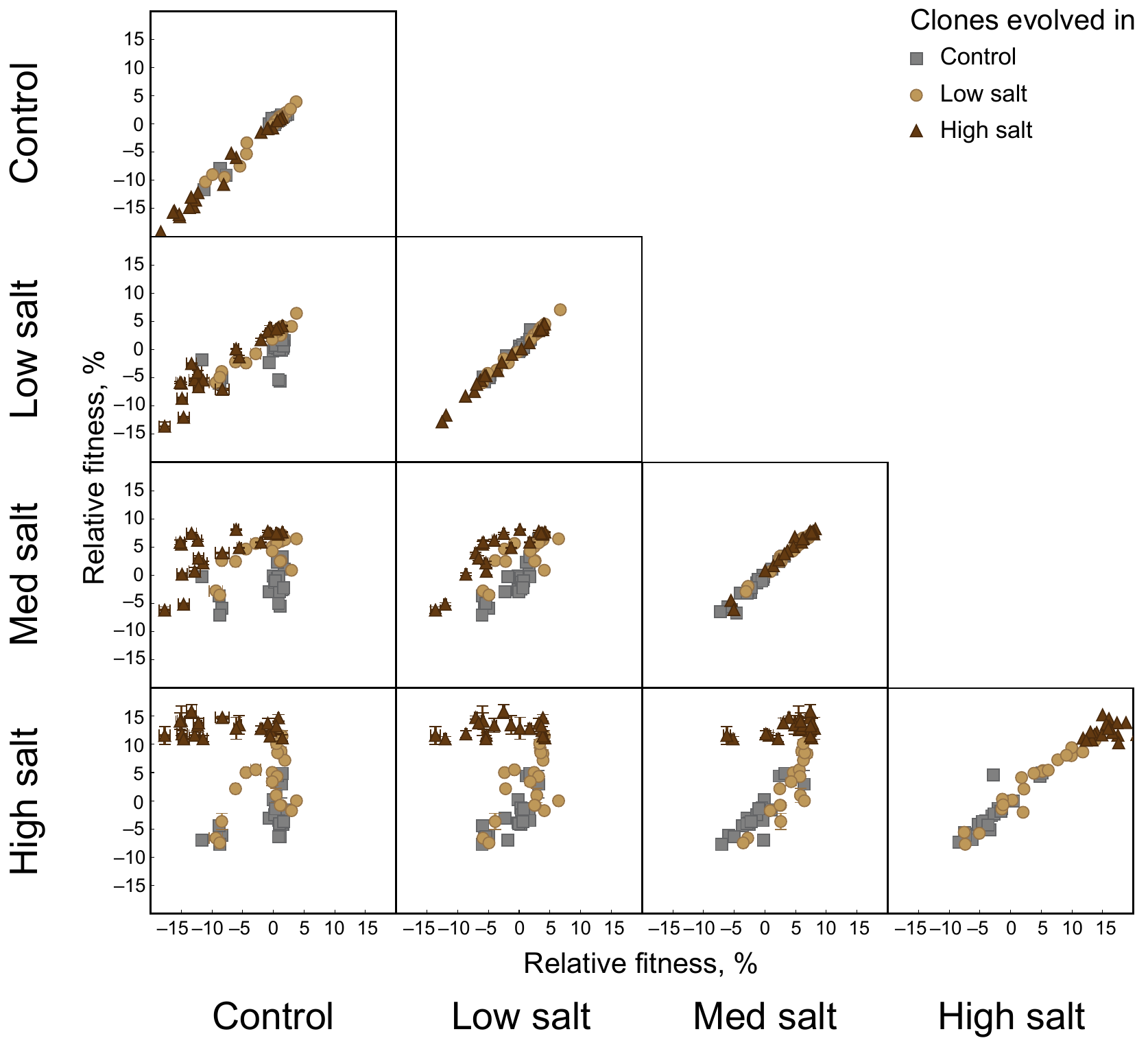}
\end{center}
\caption{{\bf Correlation between the fitnesses of clones across a range of salt concentrations. } Each symbol in a given panel represents a unique clone, where the symbol shape and color indicate the clone's home environment (see legend and Table~\ref{tableone}). Non-diagonal panels show clone fitnesses averaged over 3 replicate measurements in the environments indicated in the row and column headers (error bars denote $\pm 1$ SEM). Diagonal panels show correlations between replicate fitness measurements in the same environment. }
\label{fig: salt correlations}
\end{figure*}

\begin{figure*}[!ht]
\begin{center}
\includegraphics{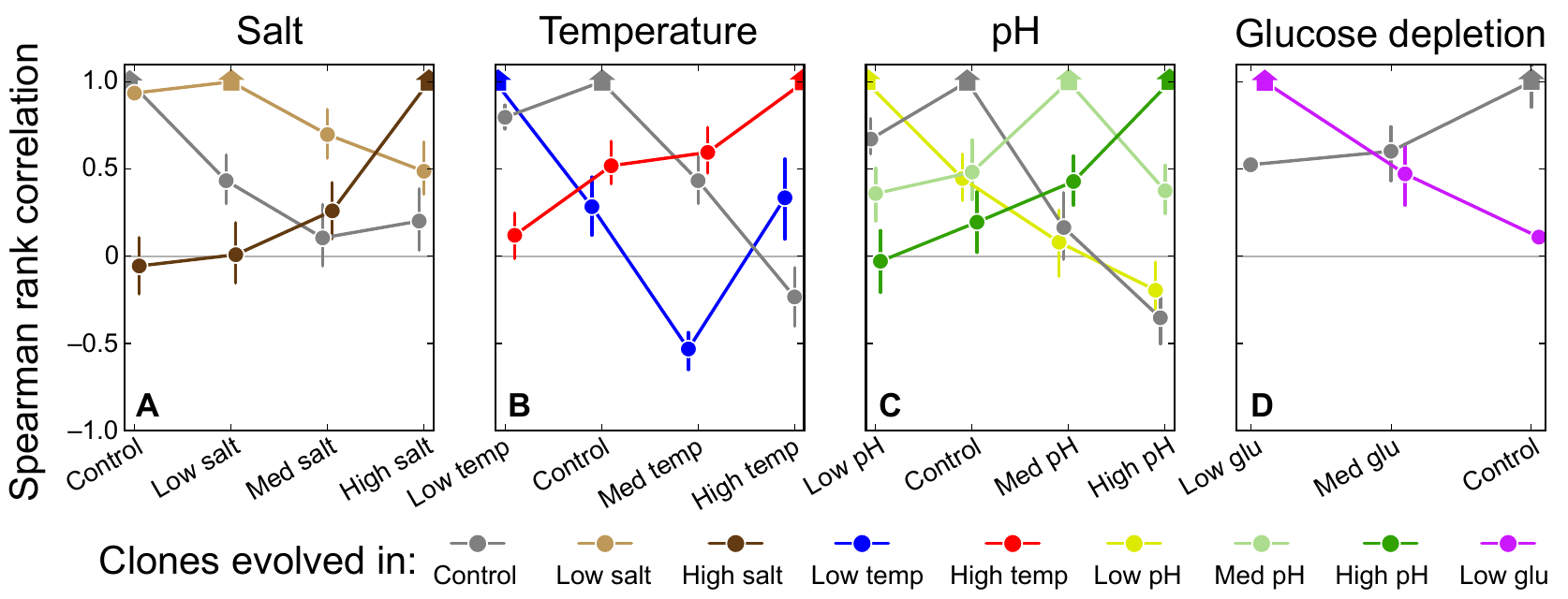}
\end{center}
\caption{{\bf Correlation coefficient between the fitness of a clone in its home environment and its fitness in other degrees of the corresponding stress.} Each symbol shows the Spearman rank correlation coefficient between the fitness of a clone measured in its home environment and in the environment indicated on the $x$-axis. These coefficients are obtained from Figures~\ref{fig: salt correlations}, S2, S3, and S4. Color correspond to each clone group's home environment (also indicated by a home symbol) as indicated in the legend and in Table~\ref{tableone}. Error bars denote the interquartile range of the error distribution for the correlation as calculated by bootstrapped resampling (see Methods). }
\label{fig: spearman all}
\end{figure*}

\section*{Results}

To analyze how adaptation to one environment leads to consequences in other similar and dissimilar conditions, we founded 180 replicate populations of budding yeast from a single common ancestor and propagated them in nine conditions (20 lines in each condition). We chose these nine conditions to span a range of intensities of four common types of continuously varying abiotic stress: temperature, pH, salt, and glucose concentration. We summarize these evolution environments in Table~\ref{tableone}. After at least 500 generations of evolution, we picked one clone per evolved population and measured its fitness in each of the different intensities of the stress in which that clone evolved. We show the complete data in Figure~S1. Details of the evolution and fitness assays are described in the Methods.

To understand how the fitness of a typical population would change across environments after adaptation to one particular condition, we calculated the median fitness of each group of twenty independently evolved clones adapted to each condition, both in the environment they adapted to (their ``home environment'') and at other intensities of the same type of stress. In Figure~\ref{fig: median fits}, we plot this median fitness of each group of clones as a function of stress intensity. As expected, all groups of clones have increased in fitness relative to the ancestor in their home environment. In addition, the clones adapted to a given environment are more fit in that environment than clones adapted to any other environment. In some cases, groups of clones that adapted in one environment also increase in fitness in similar environments (i.e. at similar intensities of that stress), indicating that the average mutation acquired in the home environment is also beneficial there. On the other hand, we typically observe tradeoffs between sufficiently distant intensities of stress, and these tradeoffs tend to be stronger in more distant environments. As a result, the fitnesses of clone groups in the most extreme conditions are ordered according to the similarity of that condition to their home environment (see Figure~\ref{fig: median fits}). For example, at the highest pH (right side of Figure~\ref{fig: median fits}B), the clones evolved at this high pH are the most fit, followed by the clones evolved in progressively more acidic conditions.

Additionally, the trade-offs we observed were consistently asymmetric from one extreme of a range of stress to the other. The group evolved in the more stressful condition (e.g. the high salt-adapted lines, dark brown in Figure~\ref{fig: median fits}A) lost more fitness in the permissive condition than vice versa (e.g. the control lines, grey in Figure~\ref{fig: median fits}A).

We note two exceptions to these general observations. First, typical trade-offs at different glucose concentrations were small to nonexistent. The group of clones evolved in low glucose showed no net fitness gain or loss in high glucose, and the group of clones evolved in high glucose improved in fitness slightly in low glucose, although not as much as the populations evolved there (Figure~\ref{fig: median fits}D). Second, clones adapted to the lowest temperature also improved in fitness at the highest temperature, although they showed a fitness decline at an intermediate temperature (Figure~\ref{fig: median fits}B). The reason for this idiosyncrasy is unclear.

To investigate how the rank order of fitnesses of clones changes across environments, we next analyzed the fitness of individual evolved clones across multiple conditions. Each panel of Figure~\ref{fig: salt correlations} shows the fitnesses of each salt-evolved clone in a pair of salt concentrations. As expected, in similar environments there is a strong correlation between fitnesses of clones regardless of their home environment. This correlation decays when we consider more different environments (i.e. looking across a row or down a column in Figure~\ref{fig: salt correlations}). In distant environments, the evolutionary history of each clone determines the structure of the correlations: each group of clones has a small variance of fitness in the home environment but a much broader range in more distant conditions. We present the analogous results for other stresses in Figures S2, S3, and S4.

We summarize the Spearman rank correlation between the fitness of each clone in its home environment and its fitness in other intensities of that stress in Figure~\ref{fig: spearman all}. These are the most important correlations from an evolutionary perspective, as they reflect how the rank order of fitness within a population would change after an environmental shift. For all types of stress, these correlations are typically positive in environments close to the home environments, and decay towards zero as the environment becomes more distant (note however the exception for lines evolved at low temperature). However, these correlations only rarely become negative, even in pairs of environments in which there are strong average fitness tradeoffs (compare the corresponding panels in Figures \ref{fig: median fits} and \ref{fig: spearman all}). Conversely, even in the absence of average fitness tradeoffs (e.g. across glucose concentrations), the correlation structure can vary substantially across environments. Finally, we note that there are some asymmetries in the correlation structure (e.g. the fitnesses of lines evolved in low salt are quite well-correlated with their fitnesses in the control environment, but the converse is not true).

\section*{Discussion}

In principle, we can interpret our observations within the classical framework that distinguishes between two models of trade-offs: mutation accumulation (MA) and antagonistic pleiotropy (AP) \citep{rose1980test, cooper2000population, MacLeanBellRainey2004}. In the MA model, accumulation of neutral mutations during evolution in one environment leads to fitness loss in another condition; these mutations lead to the deterioration of functions required in the second environment but not the first. In contrast, in the AP model, the fitness costs in a novel environment are caused by the same mutations that are adaptive in the evolution environment: there is a direct trade-off between traits.

In both the MA and AP models, we might expect that a population typically loses more fitness in conditions that are less similar to the home environment, as we observe (Figure~\ref{fig: median fits}). However, the AP model would predict that the fitnesses of evolved clones in novel environments are negatively correlated with their fitness in the home environments. By contrast, in the MA model the fitness of an individual in two environments is determined by different sets of mutations, so we expect that the fitnesses of evolved clones in a novel environment would be poorly but not negatively correlated with their fitnesses in the home environment, as we observe (Figure~\ref{fig: salt correlations}).

Although these aspects of our data are consistent with the MA picture, explaining some of our other observations in this framework would require further assumptions. For example, we consistently observe an asymmetry: populations evolved in stressful conditions lose considerably more fitness in permissive conditions than vice versa (Figure~\ref{fig: median fits}). This observation would be consistent with MA if neutral mutation rates in stressful conditions were higher than in permissive conditions. This may be the case for osmotic stress \citep{Shor2013}, but we have no reason to expect such a difference in mutation rates arising from other types of stress. Alternatively, this asymmetry could arise if mutations that are neutral in stressful conditions were highly deleterious in permissive conditions, but not vice versa --- a possible but unlikely scenario.

While our data is not inconsistent with the MA model, the dichotomy between mutation accumulation and antagonistic pleiotropy fails to capture the full spectrum of possible evolutionary scenarios. In particular, this framework relies on a binary classification of mutations into beneficial or neutral (in the home environment), while in reality each mutation can have a fitness effect of a particular size ranging from highly deleterious to highly beneficial. Likewise, each mutation also has particular effects on fitness in other environments.

The fact that mutational effects form a continuous rather than a binary set leads to important consequences for evolutionary dynamics in a constant environment. A natural way to characterize this continuous spectrum of possible fitness effects, and the evolutionary dynamics that emerge, is to use the distribution of fitness effects of new mutations (the DFE) \citep{Orr2003, EyreWalker2007, MartinLenormand2008}. To generalize this idea to multiple environments, we introduce a multi-dimensional distribution of fitness effects, which accounts for the fact that in addition to having a fitness effect in the home environment, each mutation also has effects on fitness in other environments. We refer to this as the {\em joint} distribution of fitness effects of new mutations (JDFE). The JDFE describes the probability that a new mutation has particular fitness effects in some set of environments of interest. For example, if we are interested in the effects of a mutation in two environments, we write $\rho(s_1, s_2)$ for the probability that a new mutation has fitness effect $s_1$ in environment $1$ and fitness effect $s_2$ in environment $2$.

\begin{figure*}[!ht]
\begin{center}
\includegraphics{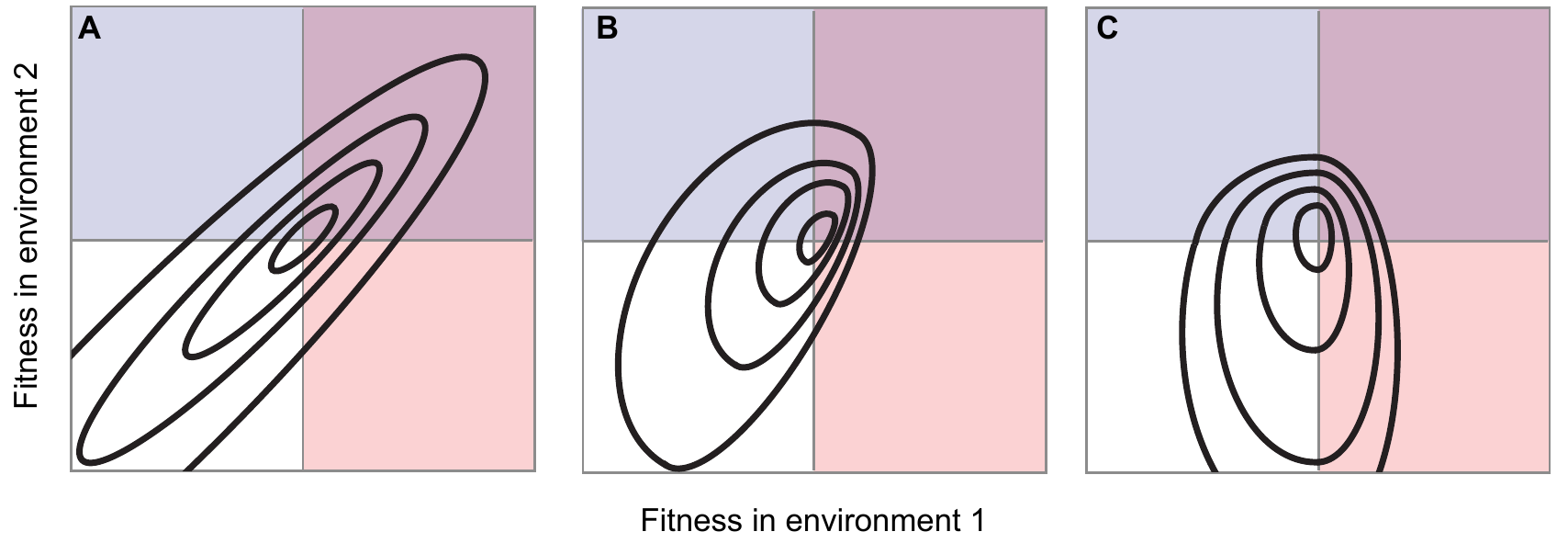}
\end{center}
\caption{{\bf Schematic illustrations of several possible joint distributions of fitness effects of new mutations (JDFEs) across two environments.} Lines show the contours of each two-dimensional JDFE. {\bf (A)} A highly correlated JDFE that might emerge if environment 2 is physically very similar to environment 1. {\bf (B)} A weakly correlated JDFE that could emerge as environments 1 and 2 become more dissimilar. {\bf (C)} An uncorrelated JDFE that might emerge as two environments become even more dissimilar. In all panels, mutations that are beneficial in environment 1 fall into the red area, and mutations that are beneficial in environment 2 fall into the blue area. Note that we could also imagine a negatively correlated JDFE (not shown here) if there is some unavoidable physiological tradeoff between fitness in two environments. }
\label{fig:JDFE}
\end{figure*}

The JDFE helps us understand and visualize the process of adaptation in one environment and its consequences in other environments. We show several examples in Figure~\ref{fig:JDFE}. Consider a population evolving in environment 1 on one of the JDFEs depicted in Figure~\ref{fig:JDFE}. Such a population would preferentially accumulate mutations that confer a fitness benefit in this environment (red-shaded regions). Thus, each clone picked from such a population would have a number of mutations sampled in this biased way from the JDFE. The patterns of fitness gain and loss that we observe in our experiment emerged from such a process, with an unknown underlying JDFE.

Two main features of the JDFE affect the evolutionary outcomes we measured. The average fitness effect in a novel environment of a mutation that is beneficial or neutral in the home environment determines whether adaptation to the home environment typically incurs a fitness loss or gain in the novel environment. We refer to this as the ``skew'' of the JDFE. For example, for a population adapting to environment 1 with the JDFE depicted in Figure~\ref{fig:JDFE}B, the mean fitness effect in environment 2 of a mutation that is beneficial in environment 1 is positive. Thus, by adapting to environment 1, a population would also gain fitness in environment 2. However, the converse is not true: the mean fitness effect in environment 1 of a mutation that is beneficial in environment 2 is negative (blue region in Figure~\ref{fig:JDFE}B), so a population adapting to environment 2 will lose fitness in environment 1. Thus, an asymmetric JDFE naturally leads to asymmetries in the patterns of typical fitness gains and losses, without invoking any additional assumptions. In fact, it is natural to expect that the JDFE between a permissive and a stressful environment would be asymmetric. In our experiment, the permissive environment is more similar to standard yeast growth media to which the organism has previously adapted, and hence it is very plausible that the supply of beneficial mutation in the permissive environment is smaller than in the stressful one. Thus, we would expect an asymmetric JDFE such as JDFEs depicted in Figure~\ref{fig:JDFE}B and Figure~\ref{fig:JDFE}C, where environment 1 is permissive and environment 2 is stressful.

The second important feature of the JDFE is the correlation between the effects of mutations in two environments, particularly among those mutations that are beneficial or neutral in the home environment. This correlation contributes to the rank correlation between fitnesses of sampled clones across environments. If all clones in our experiment acquired the same number of mutations during evolution, the correlation structure of the JDFE would exactly predict the correlation between sampled clones that we measure (Figure \ref{fig: salt correlations}). However, since different clones acquired different number of mutations, the measured rank correlation is a complex function of the overall shape of the JDFE.

The skew and correlation structure of the JDFE depend on the physical similarity between environments. Two very similar environments must have a highly correlated JDFE with a small skew, such as the one depicted in Figure~\ref{fig:JDFE}A. As the environments become less similar (for example, along a physical axis such as temperature), we expect the resulting JDFE to become less correlated and to have a larger skew (such as those depicted in Figures~\ref{fig:JDFE}B,C). When there are unavoidable physiological tradeoffs between fitness in different environments, we might expect a JDFE with a strong negative correlation.

Our measurements of fitness gain and loss across increasingly dissimilar environments are consistent with JDFEs ranging from strongly positively correlated to uncorrelated, such as those depicted in Figure~\ref{fig:JDFE}. First, we found that adaptation to one environment typically confers fitness gains in other physically similar environments (Figure~\ref{fig: median fits}); the correlations between clone fitnesses in these similar environments were positive but significantly less than one. Both observations are consistent with a JDFE that is strongly but not perfectly correlated (e.g. Figure~\ref{fig:JDFE}A). By contrast, adaptation to a given environment leads on average to trade-offs in fitness in more dissimilar environments, and in some cases these trade-offs are asymmetric. As discussed above, asymmetries in the JDFE naturally lead to observed asymmetries in the patterns of typical fitness loss.

We also found that the range of clone fitnesses was narrowest in the home environment and increased as the environment became more dissimilar. This is consistent with the picture of a highly correlated JDFE that progressively becomes less correlated as we consider more physically dissimilar environments. Typically, the mutations fixed during adaptation have a characteristic narrow range of fitness effects \citep{Good2012}, which must correspond to a broader spread of fitnesses in the other environment, even if the JDFE is strongly correlated. Moreover, as the other environment becomes more dissimilar, the correlation in the JDFE decreases, leading to a wider range of clone fitnesses in this environment.

Finally, we observed that the correlations between clone fitnesses in two environments are positive when the environments are physically similar and tend to decay towards zero as the environments become more dissimilar. Clearly, a strongly positively correlated JDFE (Figure~\ref{fig:JDFE}A) will produce a positive correlation between fitnesses of clones in the respective environments. Somewhat counterintuitively, a weakly positively correlated but skewed JDFE (such as the one in Figure~\ref{fig:JDFE}B) can produce a negative correlation, for example if the variance in clone fitnesses is primarily caused by variation in the number of mutations per clone, and these mutations tend to be deleterious in the other environment. This situation is plausible given the observation of a large variance in the number of mutations between clones sampled from adapting populations \citep{Tenaillon2012, Kryazhimskiy2014}.

\subsection*{Conclusions}
The JDFE is a natural way of describing the consequences of adaptation across multiple environments in a more detailed way than the traditional binary distinction between mutation accumulation and antagonistic pleiotropy. Our data would typically be interpreted as evidence in favor of mutation accumulation. However, our observations do not necessarily reflect mutations that are neutral in the home environment and deleterious in others. Instead, they suggest that physically similar environments give rise to a highly correlated and unskewed JDFE, which becomes progressively less correlated and asymmetric as the environments become more dissimilar. Analogously, the observation of numerous mutations displaying antagonistic pleiotropy (e.g. in mutant screens) is also consistent with positively correlated but skewed JDFEs, where most (but not all) mutations that are beneficial in one environment are deleterious in the other. Thus extensive antagonistic pleiotropy at the level of individual mutations does not necessarily imply that there are unavoidable tradeoffs between two environments, in the sense of a negatively correlated JDFE.

Importantly, even if the JDFE between two conditions is positively correlated, the evolution in an environment that fluctuates between these conditions is non-trivial. The specific details of the JDFE as well as the statistics of the environmental fluctuations would determine the precise signatures of these fluctuations in the evolution and diversity of the population. Further exploration of this link, and more detailed measurements of relevant JDFEs, are therefore important in efforts to understand evolution in fluctuating environments.

\section*{Acknowledgments}

We thank Andrew Murray, Greg Lang, Benjamin Good, and Dan Rice for useful discussions. 
This work was supported by a National Science Foundation Graduate Research Fellowship (ERJ), Burroughs Wellcome Fund Career Award at Scientific Interface (SK), the James S. McDonnell Foundation, the Alfred P. Sloan Foundation, the Harvard Milton Fund, grant PHY 1313638 from the NSF, and grant GM104239 from the NIH.

%\section*{References}
% The bibtex filename
%\bibliographystyle{cbe}
%\bibliography{refs_pleiotropy}

\onecolumngrid

\setcounter{figure}{0}
\renewcommand{\thefigure}{S\arabic{figure}}
\newpage
\section*{Supplementary Figures}
\begin{figure*}[!ht]
\begin{center}
\includegraphics[width= 0.8\textwidth]{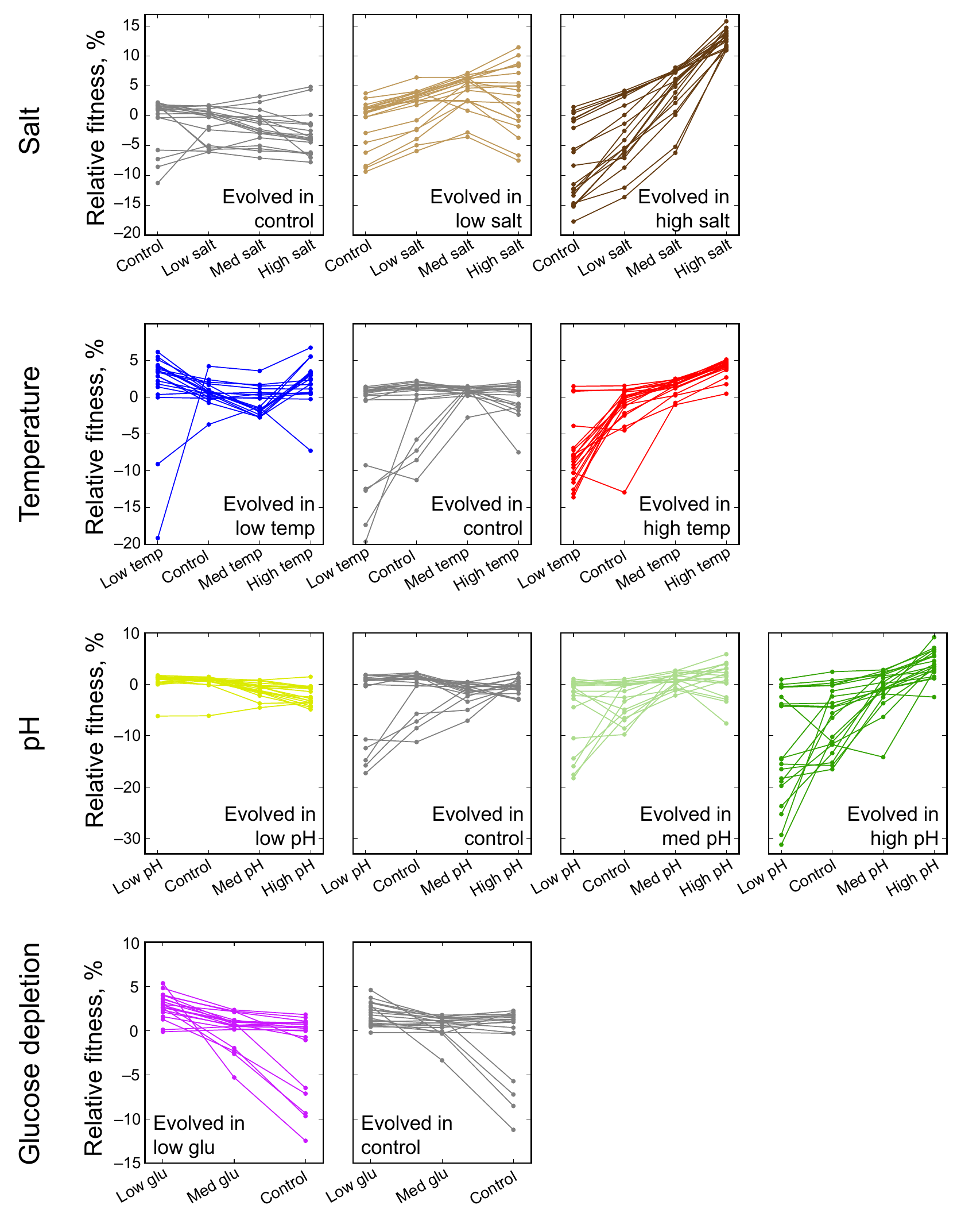}
\end{center}
\caption{{\bf Fitnesses of each clone across environments.} Each point shows the fitness of a clone in the environment indicated on the $x$-axis. Lines connect the fitnesses of the same clone across environments. Each panel shows data for a group of clones evolved in the same stressful environment; colors representing the home environments are the same as in the main text. Rows correspond to types of stress, as indicated.}
\label{fig: all data}
\end{figure*}

\begin{figure*}[!ht]
\begin{center}
\includegraphics{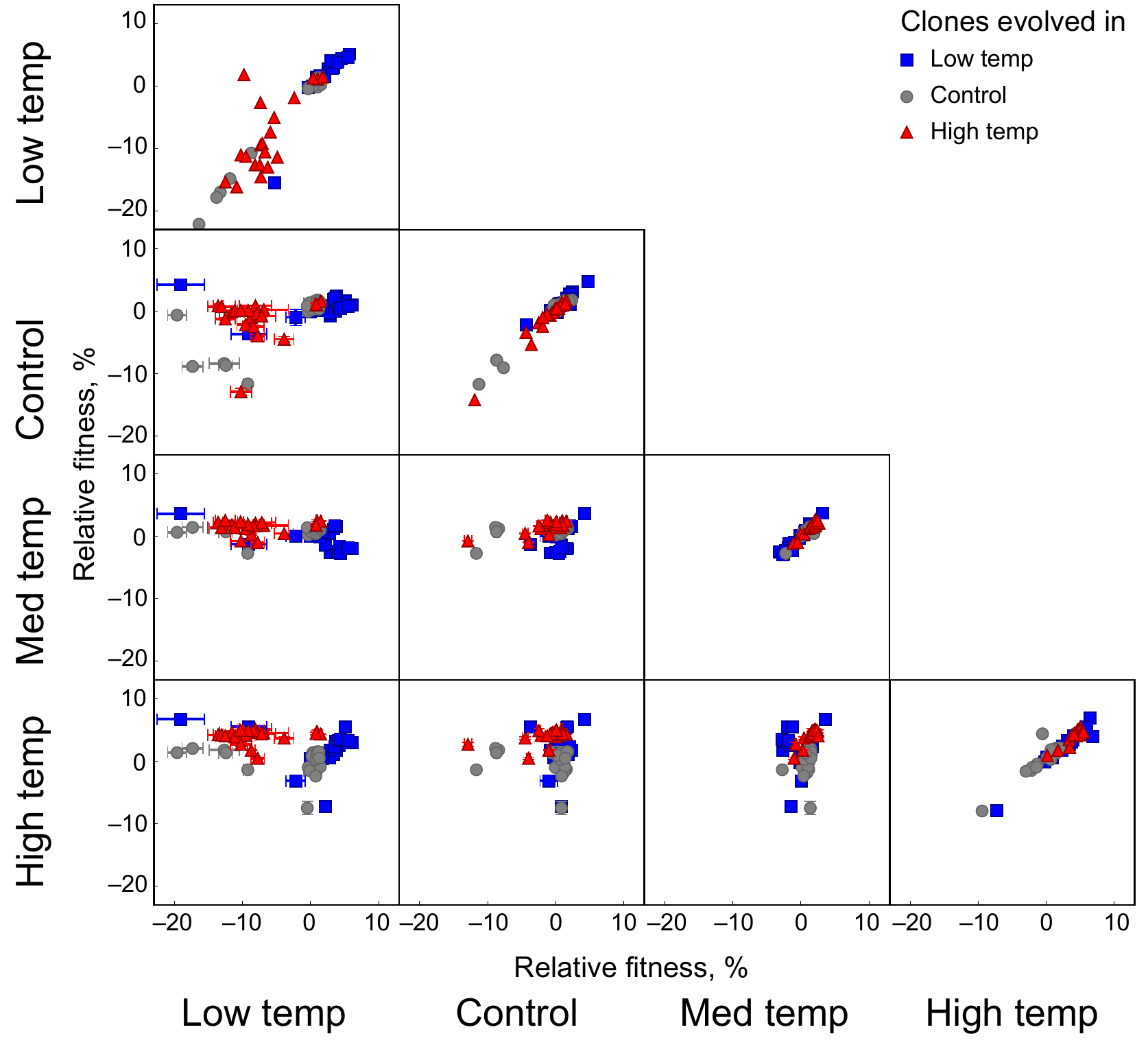}
\end{center}
\caption{{\bf Correlation between the fitnesses of clones across a range of different temperatures. } Each symbol in a given panel represents a unique clone, where the symbol shape and color indicate the clone's home environment (see legend and Table~\ref{tableone}). Non-diagonal panels show clone fitnesses averaged over 3 replicate measurements in the environments indicated in the row and column headers (error bars denote $\pm 1$ SEM). Diagonal panels show correlations between replicate fitness measurements in the same environment. }
\label{fig: Corr Temp}
\end{figure*}

\begin{figure*}[!ht]
\begin{center}
\includegraphics{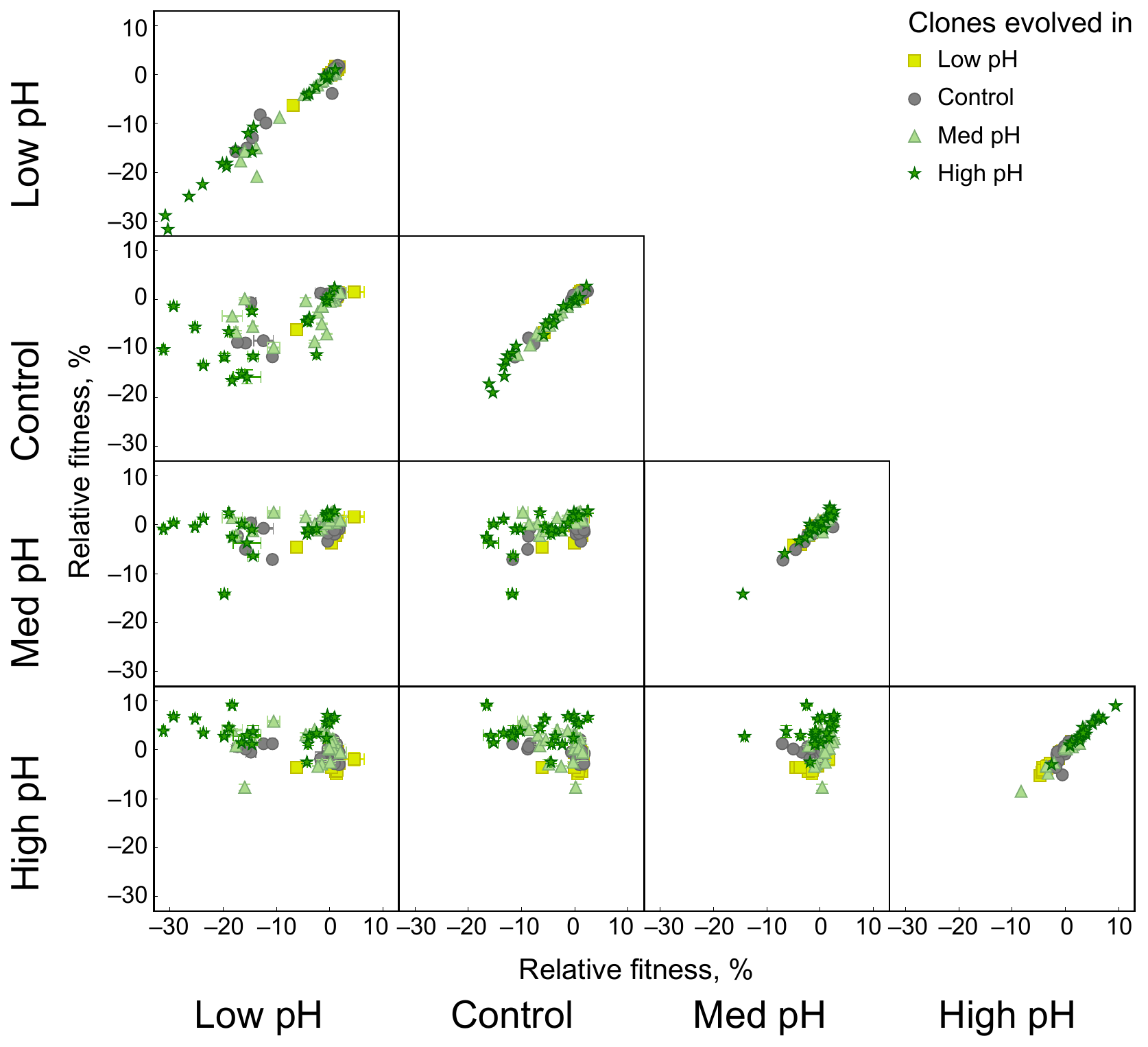}
\end{center}
\caption{{\bf Correlation between the fitnesses of clones across a range of pH. } Each symbol in a given panel represents a unique clone, where the symbol shape and color indicate the clone's home environment (see legend and Table~\ref{tableone}). Non-diagonal panels show clone fitnesses averaged over 3 replicate measurements in the environments indicated in the row and column headers (error bars denote $\pm 1$ SEM). Diagonal panels show correlations between replicate fitness measurements in the same environment. }
\label{fig: Corr pH}
\end{figure*}

\begin{figure*}[!ht]
\begin{center}
\includegraphics{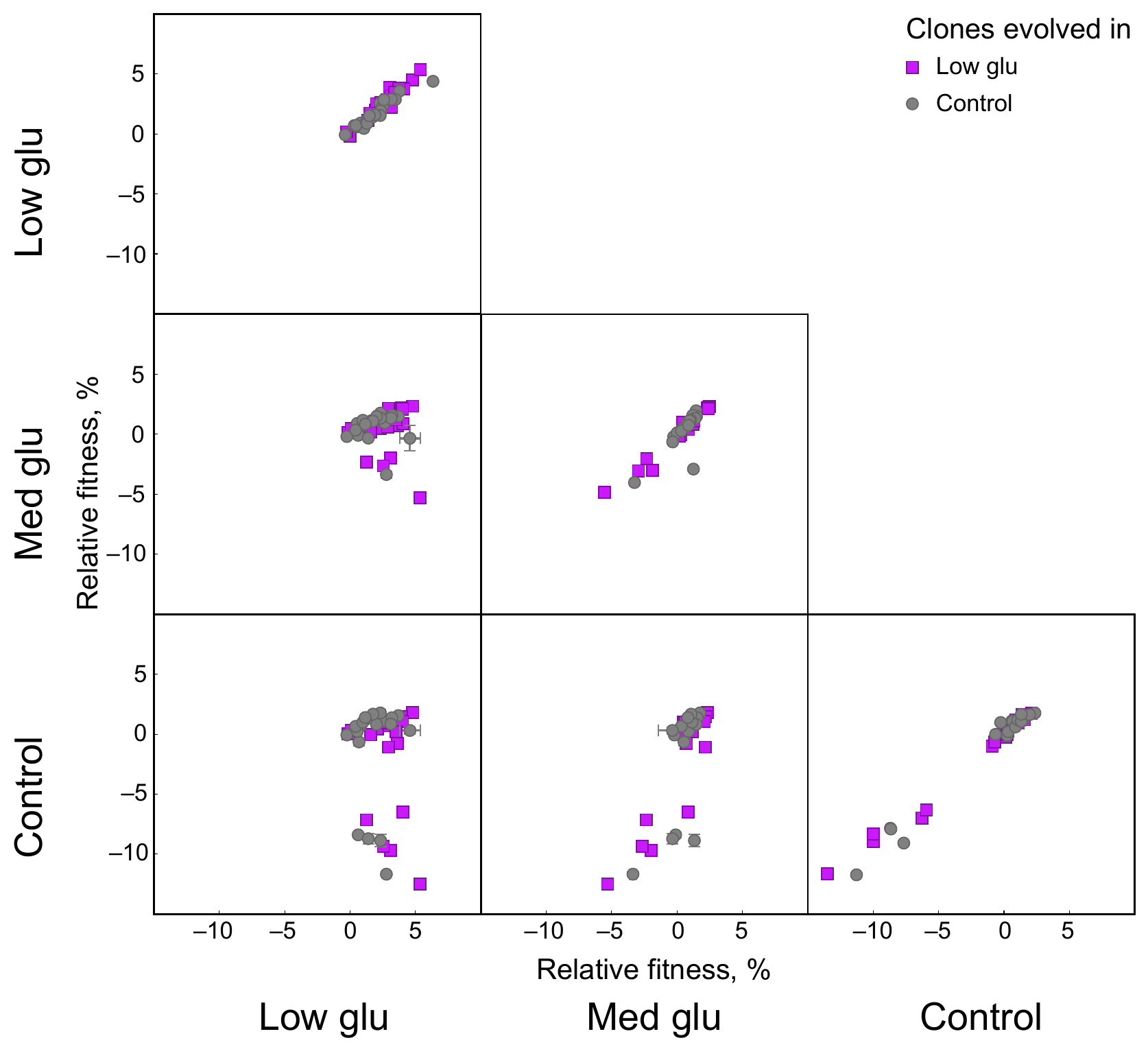}
\end{center}
\caption{{\bf Correlation between the fitnesses of clones across a range of glucose concentrations. } Each symbol in a given panel represents a unique clone, where the symbol shape and color indicate the clone's home environment (see legend and Table~\ref{tableone}). Non-diagonal panels show clone fitnesses averaged over 3 replicate measurements in the environments indicated in the row and column headers (error bars denote $\pm 1$ SEM). Diagonal panels show correlations between replicate fitness measurements in the same environment. }
\label{fig: Corr Glu}
\end{figure*}

\end{document}